\documentclass[aps,twocolumn,groupedaddress,showpacs]{revtex4}
\usepackage{ifpdf}
\usepackage{graphicx}
\usepackage{epsfig,amsbsy,bm,amssymb}
\usepackage{hyperref}
\usepackage{pslatex}
\usepackage{color}
\usepackage{todonotes}
\usepackage{marginnote}

\newcommand{\isum}%
{\mathop{\hbox{$\displaystyle\sum\kern-13.2pt\int\kern1.5pt$}}}

\renewcommand{\k}{{\bm k}}

  \newcommand{\e}{{\bm e}}

\newcommand{\bt}{\begin{tabular}}
\newcommand{\et}{\end{tabular}}

\newcommand{\eref}[1] {(\ref{#1})}
\newcommand{\Eref}[1] {Eq.~(\ref{#1})}

\newcommand{\Fref}[1] {Figure \ref{#1}}

\renewcommand{\H}{H$_2$~}

\newcommand{\br}{\begin{eqnarray*}}
\newcommand{\er}{\end{eqnarray*}}
\newcommand{\ba}{\begin{eqnarray}}
\newcommand{\ea}{\end{eqnarray}}
\newcommand{\be}{\begin{equation}}
\newcommand{\ee}{\end{equation}}

\newcommand{\bp}{\begin{minipage}}
\newcommand{\ep}{\end{minipage}}

\begin{document}
\bibliographystyle{apsrev}

\title {Instantaneous ionization rate as a functional derivative.}

\author{I. A. Ivanov$^{1}$}
\email{igorivanov@ibs.re.kr}
\author{C. Hofmann$^{2}$}
\author{L. Ortmann$^{2}$}
\author{A. S. Landsman$^{2,3}$}
\author{Chang Hee Nam$^{1,4}$}
\author{Kyung Taec Kim$^{1,4}$}
\email{kyungtaec@gist.ac.kr}

\affiliation{$^{1}$Center for Relativistic Laser Science, Institute for
Basic Science, Gwangju 61005, Korea}


\affiliation{$^2$Max Planck Institute for the Physics of Complex Systems, Noethnitzer Strasse 38, 01187 Dresden, Germany}

\affiliation{$^3$Max Planck Center for Attosecond Science/Department of Physics, Pohang University of Science and Technology, Pohang 37673, Korea}

\affiliation{$^4$Department of Physics and Photon Science, GIST, Gwangju 61005, Korea}

\date{\today}

\begin{abstract}
We describe an approach defining instantaneous ionization rate (IIR)
as a functional derivative of the total ionization probability. The definition
is based on physical quantities which are directly measurable, 
such as the total ionization probability and the waveform of an ionizing pulse. The definition is, therefore,
unambiguous and does not suffer from gauge non-invariance. We compute 
IIR by solving numerically the time-dependent Schr\"odinger equation for the hydrogen
atom in a strong laser field. In agreement with some previous results using attoclock methodology, 
the IIR we define does not show measurable delay in strong field tunnel ionization.

\end{abstract}

\maketitle

The notion of the instantaneous ionization rate (IIR) proved extremely fruitful for understanding 
physics of tunelling ionization, the ionization regime characterized by small values of the 
Keldysh parameter $\gamma=\omega/E_0\sqrt{2|I|} \lesssim 1$ \cite{Keldysh64,kri} 
(here $\omega$, $E_0$ and $I$ are the frequency, field strength and ionization potential
of a target system expressed in atomic units). 

Qualitatively, the fact that 
the IIR is a function which is sharply peaked near the local maxima of the 
electric field of the pulse, can be used to pinpoint the most probable electron trajectory.  
This provides a basis for the design and interpretation of the results of well-known experimental techniques, such as 
attosecond streaking, or angular attosecond streaking \cite{streak_camera,as0,angstr} allowing to 
follow electron dynamics at the attosecond level of precision. 
Recently, the temporarily
localized ionization at the local maxima of the laser field has been used as a fast temporal gate to measure the 
laser field \cite{optica}. 

Quantitatively, the notion of IIR 
underlies many successful simulations of tunneling ionization phenomena, relying on
classical (CTMC method) \cite{tipis} or quantum trajectories (QTMC) \cite{Li2014a,Shvetsov-Shilovski2016} Monte-Carlo simulations. These methods become
practically indispensable if the system in question is too complicated to allow an {\it ab initio} treatment 
based on the numerical solution of the time-dependent Schr\"odinger equation (TDSE). Even if numerical
solution of the TDSE is possible, these methods may provide physical insight which is not obvious from the 
TDSE wave-function. Accurate quantitative calculations using these approaches, which agree well with  the 
{\it ab initio} TDSE  have been reported 
in the literature \cite{tipis,cmtc1,cusp3,arbm,tipis_naft,landsman2015,landsman2016}. In these approaches
the quantum-mechanical Keldysh-type theories \cite{Keldysh64,Faisal73,Reiss80,ppt} 
are used to set up initial conditions for the subsequent electron motion \cite{cusp3,arbm,tipis_naft}.
IIR in the well-known ADK form \cite{adk1,adk}, or more refined Yudin-Ivanov IIR 
\cite{YI} provides a measure allowing to assign probability to an ionization event occurring at a given
moment of time inside the laser pulse duration. 

The notion of the ionization event occurring at a given time and, correspondingly, the notion of IIR 
are not free from certain ambiguity, however. An interpretation of the ionization event which is
often used is based on the  imaginary 
time method \cite{ppt,tunr2}. In this picture an electron enters the tunelling barrier at some 
complex moment of time with complex velocity. Upon descending on the real axis, the velocity and 
coordinates corresponding to the most probable electron trajectory become real, which 
can be interpreted as the electron's exit from under the barrier. This picture, however, 
cannot be taken unreservedly, since the path which descends on the real axis is not unique and
can be deformed, in principle, to cross the real time-axis at almost any given point \cite{tunr2}. 
An approach allowing to define IIR from the solution of the 
TDSE has been described recently \cite{inst1}. 
In this approach the IIR is defined by projecting out contributions of the bound states from the solution of the TDSE.
The authors found that the IIR thus defined lags behind the local maxima of the electric field, which 
suggests a non-zero tunneling time. A shortcoming of this definition of the IIR, however, is its
non-gauge-invariant character. The projection of the solution of the TDSE on the subspace of the bound states performed during the interval of the pulse duration depends generally on the gauge used to describe atom-field interaction. Another approach allowing to
define IIR from the solution of the TDSE based on the notion of the electron flux was given in \cite{Teeny2016}.

In the present work we describe a novel approach to IIR, which is based on the 
notion of a functional derivative. 
Total ionization probability can be considered as a functional $P[E]$ of the waveform $E(t)$, which
maps the electric field of the pulse into a real number. For the regime of the tunneling ionization
we are interested in below, this functional is highly non-linear and 
cannot be described in a closed form. A simplification is possible if we consider a waveform which can be
represented as $E(t)= E_f(t)+\delta E(t)$, with fundamental field $E_f(t)$ and signal field $\delta E(t)$.
To be more specific, let us assume that the fundamental field is linearly polarized (along $z-$ axis) 
and is defined by the vector potential ${\bm A_f(t)}$:

\begin{equation}
{\bm A_f(t)}= -\hat {\bm z} {E_0\over \omega}\sin^2{\left\{\pi t\over T_1\right\}}\sin{\omega t} \ ,
\label{ef}
\end{equation}

with peak field strength  $E_0$, carrier frequency $\omega$, and total duration $T_1=NT$, where 
$T=2\pi/\omega$ is an optical period (o.c.) corresponding to the carrier frequency $\omega$, with $N \in \mathbb{N}$.

We assume the signal field to vanish outside the interval $(0,T_1)$ of the fundamental pulse
duration. If the signal field $\delta E(t)$ is sufficiently weak we can write:

\be
\delta P= P[E_f+\delta E]-P[E_f] \approx \int\limits_0^{T_1} {\delta P\over \delta E_f(t)} \delta E(t) dt\ ,
\label{fd}
\ee

where $\displaystyle {\delta P\over \delta E_f(t)}$ is a functional derivative of the 
functional $P[E]$ evaluated for $E(t)= E_f(t)$.

On the other hand, using the customary definition of IIR, we can write for the 
probability of ionization driven by the combined field $E(t)=E_f(t)+ \delta E(t)$:

\be
P_{\rm inst}= \int\limits_0^{T_1} W_{\rm inst}(E(t))\ dt \ ,
\label{in1}
\ee

where $W_{\rm inst}(E(t))$ is the IIR, which by definition depends on the 
instantaneous value $E(t)$ of the electric field. We introduced the notation $P_{\rm inst}$ in \Eref{in1} to emphasize that 
this expression pertains to the notion of the IIR.
\Eref{in1} gives a very particular case of the functional $P[E]$ -- clearly not every functional can be
represented in this way. 

From \Eref{in1} we obtain the change of the ionization probability $P_{\rm inst}$ 
due to the presence of the 
weak signal field:

\be
\delta P_{\rm inst} \approx \int\limits_0^{T_1} {d W_{\rm inst}(E_f(t))\over dE(t)} \delta E(t) dt\ ,
\label{fd1}
\ee

Taking into account that $\delta E(t)$ in \Eref{fd} and \Eref{fd1} is arbitrary (provided it is small and
vanishes outside the interval $(0,T_1)$), one can see that, if the treatment based on the notion of the IIR
is justified (i.e. using \Eref{in1} with IIR $W_{\rm inst}(E(t))$
depending only on the instantaneous value of the electric field one obtains accurate value for the 
total ionization probability), one must have:

\be
{\delta P\over \delta E_f(t)} \approx {d W_{\rm inst}(E_f(t))\over dE(t)}\ .
\label{fd2}
\ee

We note that the quantity on the r.h.s of \eref{fd2} is an ordinary derivative which depends on time only through
the instantaneous value of the electric field. The quantity on the l.h.s., on the other hand, is a 
functional derivative, which depends on time (through the time moment $t$ at which differentiation is
performed), and on the waveform $E(t)$, i.e., on the complete history, past and future, of the pulse. 
Suppose, for instance, that we represent the waveform
\eref{ef} as a Fourier series:

\be
E(t)= \sum\limits_m a_m e^{im \Omega t} \ ,
\label{fou}
\ee

where $\Omega=2\pi/T_1$. In general, the functional derivative on the l.h.s. of \Eref{fd2} depends on the 
whole set ${a_m}$, while the ordinary derivative on the r.h.s. is a function of the particular combination of these 
coefficients.  To express this differently, suppose, 
that we fix the pulse shape in \Eref{ef} and allow only the field amplitude $E_0$ to vary. The l.h.s of \Eref{fd2}
would then generally be a function of two variables $E_0$ and $t$, while the r.h.s. would depend only on a particular 
combination of $E_0$ and $t$ which gives the instantaneous field strength $E(t)$. 
Exact equality in 
\Eref{fd2}, therefore, cannot be achieved in general. This clearly demonstrates the approximative character of the notion of the 
IIR. The more accurate expression \eref{fd},
based on the functional derivative of the 
ionization probability functional, depends not only on time, but on additional variables describing the waveform as well. 
For brevity, we will dub below the functional derivative in \Eref{fd} an "exact ionization rate", though as 
one can see from \Eref{fd2}, it rather corresponds to the derivative of the IIR with respect to 
the electric field. 

To compute the exact ionization rate in \Eref{fd2} for a given moment of time $\tau$ and 
a given waveform $E_f(t)$, one can calculate numerically the modulation of the ionization probability $\delta P$ using 
$E_f(t)$ as a fundamental field and employing a special form $\delta E(t,\tau)= \alpha \delta(t-\tau)$ 
of the signal field, containing the Dirac delta-function.
We did this by solving numerically the TDSE for a hydrogen atom in the presence of the 
pulse \eref{ef}. We considered pulses with the low carrier frequency of $\omega=0.02$ a.u. We need to work with low frequencies
to stay within the framework of the adiabatic theory, which makes the notion of the instantaneous ionization 
rate at least qualitatively applicable. 
We report below
results of the calculations with total pulse durations $T_1$ of one or two optical cycles respectively, and various 
peak field strengths $E_0$ (chosen such as to remain in the tunelling regime of ionization). 

The solution of the TDSE has been found by representing the wave function as a series in 
spherical harmonic functions, and discretizing the resulting system of radial equations 
on a grid with the step-size $\delta r=0.05$ a.u. in a box of the size $R_{\rm max}=700$ a.u., which was sufficient for the 
pulses of a short duration we consider below. 
More details about the numerical procedure used to solve the TDSE can be found in Refs. \cite{cuspm,circ6}.

Total ionization probability, which we need for the practical implementation of the definition \eref{fd2}, 
is found by decomposing the wave function at the end of the pulse as:

\be
\Psi(T_1)= \phi + \chi  ,
\label{dec}
\ee

where $\phi= \hat Q \Psi(T_1) $, $\chi= (\hat I- \hat Q) \Psi(T_1)$ , and $\hat Q$ is the projection operator on the
subspace of the bound states of the field-free atomic Hamiltonian. Total ionization probability $P$ can
be found then as the squared norm $P=||\chi||^2$. Projection operator $\hat Q$ is obtained by computing numerically (employing the same grid
we used to solve the TDSE) the bound states $|nl\rangle$ of the field-free Hamiltonian:

\be
\hat Q= \sum\limits_{nl\atop l=0}^{L_b} |nl\rangle\langle nl| \ ,
\label{qq}
\ee

where we retain all eigenvectors we obtain  $0\le l \le L_b$. We used $L_b=12$ in the calculations below
(having checked that results are well converged with respect to this parameter). We note that definition of the ionization
probability as the squared norm of $\chi$ is formally equivalent to the often used definition in terms of the projection on
the asymptotic scattering states $\phi_{\k}$: $\displaystyle P= \int d\k |\langle \phi_{\k}|\Psi(T_1)\rangle|^2$ (assuming 
normalization $\langle \phi_{\k'}|\phi_{\k}\rangle= \delta(\k-\k')$).
Indeed, substituting decomposition \eref{dec} in this equation, and using orthogonality of 
$\chi$ to the subspace of the bound states, we obtain again $P=||\chi||^2$. In practice, the prescription 
$P=||\chi||^2$, we use in the calculations below, is preferable. The reason for this is that calculation of the total ionization probability 
by projecting $\Psi(T_1)$ on the set of the scattering states implicitly presumes the strict orthogonality of the 
scattering and bound atomic states. We are interested below not in the ionization probability itself, but rather 
in its relatively small variations due to the weak signal field. Even small non-orthogonality of the 
scattering and bound states unavoidable in numerical calculations may lead to a significant loss of precision.

The definition of exact ionization rate in \Eref{fd} is based on the physically
observable quantities:  the electric field of the pulse and modulation of the total ionization probability. 
It is clearly gauge invariant. It is immaterial, therefore, which gauge describing the atom-field interaction is used 
when solving the TDSE (provided, of course, that a sufficient
level of numerical accuracy has been achieved). We used the length (L-) gauge. Convergence of the expansions
in spherical harmonic functions we use to represent the solution of the TDSE was achieved by 
including spherical harmonic functions of the rank up to $L_{\rm max}= 70$. In numerical calculations we have to regularize, 
of course, the  expression $\delta E(t,\tau)= \alpha \delta(t-\tau)$ for the 
signal field. We used the regularization:

\be
\delta E(t,\tau)= {\alpha\over \epsilon} \exp \left\{ -{(t-\tau)^2\over\epsilon^2} \right\} \ ,
\label{reg}
\ee

with $\epsilon=T/1000$ (here $T$ is an optical period corresponding to the carrier frequency we use), and
$\alpha=0.001$. The value of $\alpha$ is to be chosen so that the higher order 
functional derivatives in the Taylor expansion for the the ionization probability functional in \Eref{fd} can be omitted. 
That our choice of this parameter warrants such an omission is shown in the Supplemental Material, which presents a
detailed study of the influence of the parameters $\alpha$ and $\epsilon$ on the accuracy of the calculation.
It is worth noting that even for $\epsilon=T/30$ the relative error we get for the ionization
probability is of the order of one percent. FWHM of the pulse \eref{reg} for this value of $\epsilon$ is about 
$300$ attoseconds. Delta-function-like pulses of such duration have already been produced in the laboratory \cite{delta}, 
which makes possible experimental measurements of the IIR which we define in the present work.

We will fix below the functional form of the fundamental pulse shape in \Eref{ef} and allow only
the peak field strength $E_0$ of the fundamental field to vary. In figures below we show and analyze 
not the functional derivative itself but a proportional quantity: the 
first variation of the ionization probability $\delta P(E_0,\tau)$, obtained if we substitute expression \eref{reg}
for the signal field in \Eref{fd}.

\begin{figure}[h]
\includegraphics[width=0.8\linewidth]{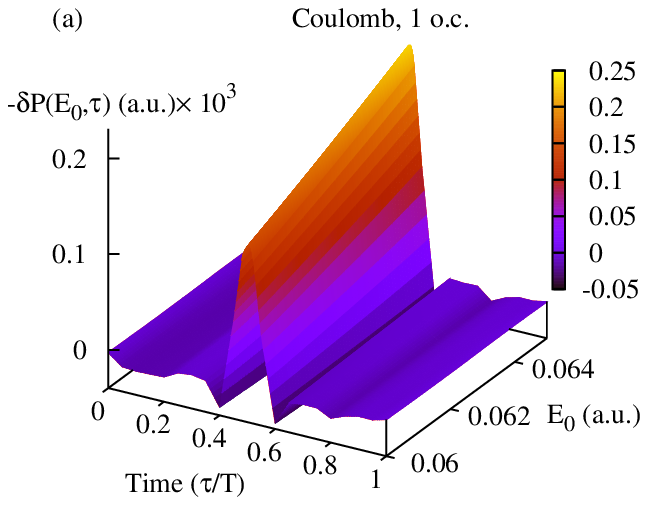}
\includegraphics[width=0.8\linewidth]{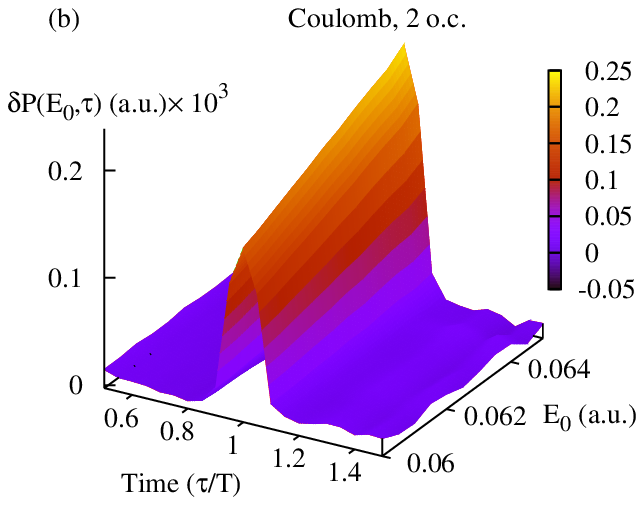}
\caption{(Color online) First variation of the ionization probability $\delta P(E_0,\tau)$ obtained from TDSE for Coulomb
potential for the total pulse duration of 1 o.c. (a) and 2.o.c (b). }
\label{fig1}

\end{figure}

We show in \Fref{fig1} the first variation $\delta P(E_0,\tau)$
computed following the numerical procedure
outlined above as a function of time $\tau$ and electric field strength $E_0$ for the pulses \eref{ef} with
total duration $T_1$ of one or two optical cycles.

\begin{figure}[h]
\includegraphics[width=0.8\linewidth]{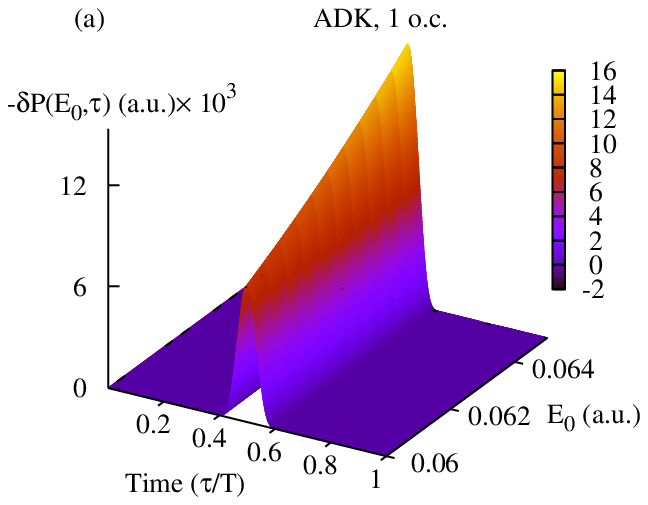}
\includegraphics[width=0.8\linewidth]{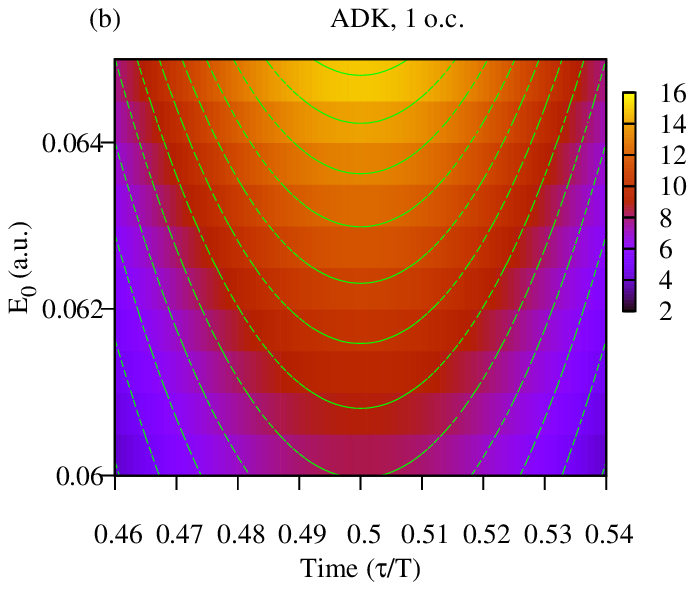}
\caption{(Color online)  First variation of the ionization probability $\delta P(E_0,\tau)$ 
obtained for ADK IIR for the total pulse duration of 1 o.c. (a). 
Contour plot of $\delta P(E_0,\tau)$ (b).
}
\label{fig2}
\end{figure}

In Fig.2a we present  $\delta P(E_0,\tau)$ obtained using the ADK 
expression \cite{adk1,adk} for the IIR.
Since the ADK ionization probability assumes an instantaneous relation between the field strength 
and the probability of ionization at any given time, this naturally leads to a definition of the IIR 
following equation \Eref{fd1}. 

More detailed picture of the IIR's emerges by looking at the 
contour plots (i.e. lines of equal elevation) of  $\delta P(E_0,\tau)$ in the $E_0,\tau$-plane. 
Of particular interest is, of course, the behavior of the IIR near the local maxima of the electric field where 
the IIR attains its largest values. We will concentrate, therefore, on the region of $\tau$-values close to the 
highest maximum of the electric field strength. 
We will consider below only the
pulses with a total duration of one o.c., the results for the pulses with duration of 2 o.c. have been found to be 
essentially the same as for 1 o.c. duration.

Since the ADK IIR is a function of the instantaneous electric field only,
lines of constant elevation in this case are just the lines satisfying $|E(\tau)|={\rm const}$. Contour lines given by this equation
are shown in Fig. 2b. 
The curves are perfectly symmetric with respect to the 
instant of time when electric field of the pulse has a maximum, and are parabolic near this point. These features,
of course, can be readily deduced from the fact that the electric field of the pulse \eref{ef} is an even function, symmetric 
about the midpoint of the pulse.

Any asymmetry of the contour lines about the midpoint of the pulse 
could be interpreted as a manifestation of the tunneling delay, as was done in 
\cite{inst1}, where a lag was found for the IIR defined in that work by projecting out bound states contributions
from the TDSE wave-function. 
Or in \cite{Teeny2016}, where the authors monitored the probability current density at the (albeit adiabatic) exit point, as calculated by a one-dimensional TDSE solution. The point of view that the tunneling delay has a non-zero value has also been expressed
in the works \cite{Teeny2016,landsman_bohm,nonzerot1,nonzerot2,nonzerot3}, as opposed to the papers
advocating view of tunelling as an instantaneous process \cite{tor,Nie,Nie1}. 
Fig. 3a shows contour lines of  $\delta P(E_0,\tau)$ obtained from TDSE calculation for hydrogen with 
the Coulomb potential. 
The absence of any appreciable asymmetry in the Figure suggests that 
our definition of the IIR gives us an essentially zero time delay for the Coulomb potential. 

We considered also the case of the Yukawa potentials  $V(r)=-A\e^{-{r\over a}}/r$ with 
different screening parameters $a$. For every $a$ we adjusted the value of $A$ 
so that the resulting ionization potential for the ground state was always 
$0.5$ a.u., corresponding to hydrogen. Results for the Yukawa potentials shown 
in Fig 3b, Fig.3c again show
an absence of any appreciable asymmetry of the contour lines, and consequently the time delay.

\begin{figure}[h]
\includegraphics[width=0.7\linewidth]{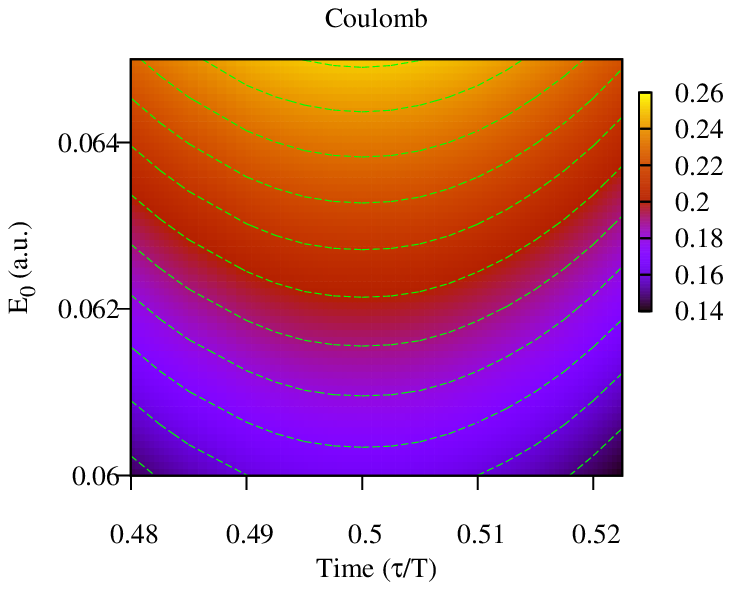}
\includegraphics[width=0.7\linewidth]{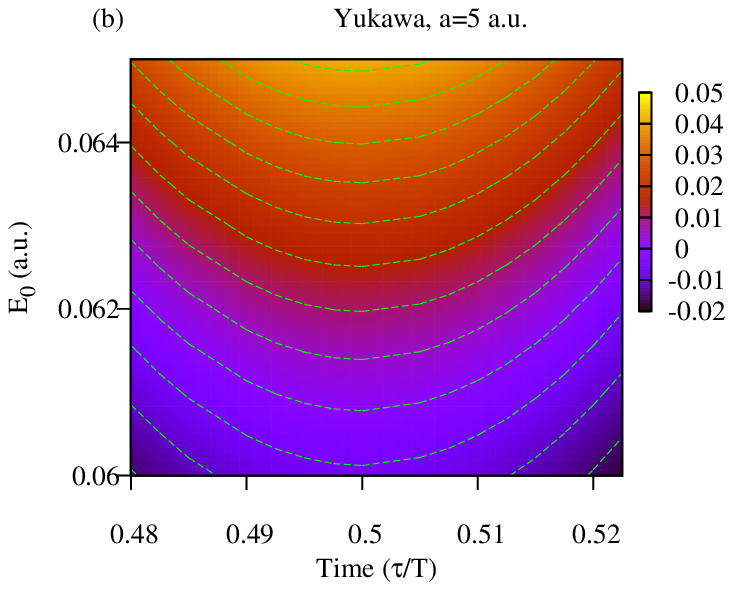}
\includegraphics[width=0.7\linewidth]{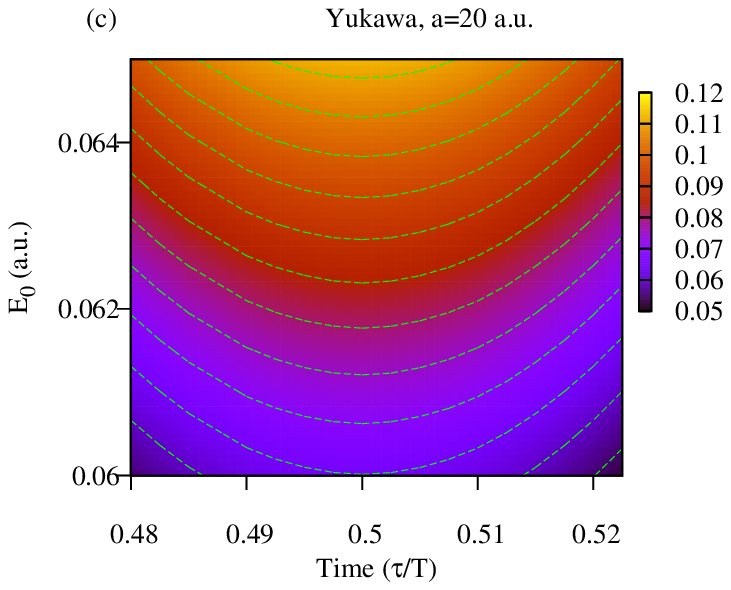}
\caption{(Color online) Contour plots of 
$\delta P(E_0,\tau)$ for  
the Coulomb potential (a) and Yukawa potentials with different screening parameters (b),(c).}
\label{fig3}
\end{figure}

To summarize, we described an approach allowing to define the IIR 
as a functional derivative of the total ionization probability. This approach provides an unambiguous
definition of the IIR. In particular, it is based on directly measurable quantities, such as the total ionization probability and
the waveform of the pulse, which makes it gauge invariant. 
We studied the IIR thus defined using tunneling ionization of systems with 
Coulomb and Yukawa potentials as examples.
In agreement with some previous results using attoclock methodology (which assume the most probable electron trajectory to begin 
tunneling at the peak of the laser field), the IIR we define does not show any appreciable 
tunelling delay in strong field ionization for the case of hydrogen.

\newpage



\end{document}